\documentclass[a4paper,11pt]{article}
\pdfoutput=1 % if your are submitting a pdflatex (i.e. if you have
\usepackage{jinstpub} % for details on the use of the package, please
\usepackage{xspace}                     
\newcommand{\tensorflow}{\ensuremath{\mathrm{TenorFlow}}\xspace}
\newcommand{\scikinC}{\ensuremath{\mathtt{scikin}}\xspace}

\usepackage{lineno}
\title{\boldmath Machine Learning for the LHCb Simulation}

\author[a]{Lucio Anderlini}

\affiliation[a]{Istituto Nazionale di Fisica Nucleare -- Sezione di Firenze\\% 
    via G. Sansone, 1 -- Sesto Fiorentino -- Italy}

\emailAdd{Lucio.Anderlini@fi.infn.it}

\abstract{
Most of the computing resources pledged to the LHCb experiment at CERN are necessary 
to the production of simulated samples used to predict resolution functions on the 
reconstructed quantities and the reconstruction and selection efficiency.
Projecting the Simulation requests to the years following the upcoming LHCb Upgrade, 
the relative computing resources would exceed the pledges by more than a factor of 2. 
In this contribution, I discuss how Machine Learning can help to speed up the 
Detector Simulation for the upcoming Runs of the LHCb experiment. 
}


\collaboration[c]{on behalf of the LHCb Collaboration}

\proceeding{
Artificial Intelligence for the Electron Ion Collider (experimental applications)\\
7-10 September, 2021}

\begin{document}
\maketitle
\flushbottom
The LHCb 
detector~\cite{Alves:2008zz,Aaij:2014jba} is a
single-arm forward spectrometer covering the pseudorapidity range $2 < \eta < 5$, designed for
the study of particles containing $b$ and $c$ quarks. 
It includes a silicon-strip vertex detector surrounding the $pp$ interaction region that 
allows $c$ and $b$ hadrons to be identified from their characteristically long flight distance;
a tracking system that provides a measurement of momentum, $p$, of charged particles. 
The Particle Identification apparatus is composed of two 
ring-imaging Cherenkov detectors that are able to discriminate between
different species of charged hadrons, two calorimeters able to discriminate electrons from 
hadrons and muon, and a set of four muon stations downstream the calorimeters. 

The size of each event acquired by the LHCb experiment is smaller than for the other 
experiments at the LHC~\cite{Boccali:2019yhm}.
LHCb aims indeed at collecting very large and pure data samples of 
$b$ and $c$ hadron decays, to reduce the statistical uncertainties to a negligible level 
and to access the rarest decays modes occurring to less than one hadron in a billion. 
Both the search for rare events and the precision measurements on heavy quarks, part 
of the mainstream physics programme of the LHCb experiment since its conception, 
put severe constraints on the accuracy and precision of the Simulation. For example, 
the uncertainties on the efficiencies determined from Simulation should not exceed the 
tiny statistical errors obtained with the large samples collected, while providing a good 
description of the rejection power of unlikely background events 
to effectively design the searches for ultra-rare decays. 

As a consequence, very large productions of simulated samples are necessary to perform 
several important data analyses. These simulations are performed with a released and frozen 
detector configuration and with beam and alignment conditions tuned to provide the best possible representation 
of each period of data taking.
It has been estimated that the vast majority of the CPU requests by the LHCb experiment has been 
necessary to produce simulated samples, and that the requests will exceed the pledged resources since 
2022~\cite{LHCB-FIGURE-2019-018}.

Important speed-up factors in the simulation can be achieved by reusing the underlying event 
associated to the production of a heavy hadron for multiple decays of the latter~\cite{Muller:2018vny}.
An alternative way forward to reduce the amount of computing resources 
needed to produce simulated samples, is to model the response of each detector element 
building a sort of cache instead of simulating the radiation-matter interaction 
for each particle in each simulated event. 

Machine Learning offers an extremely powerful technique to model the detector response. 
In particular, Neural Networks are capable of encoding for each impinging particle the 
distribution of possible responses for each detector element, retaining the capability of 
modelling the tails in the detector response giving origin to rare background events, while 
drastically reducing the CPU time necessary to simulation.

The rest of this document is organised as it follows. In Section~\ref{sec:fullsim}, I will briefly 
review the concept of Detailed Simulation to establish the baseline on which injecting 
the Generative Models for the energy depositions in the calorimeters and for higher level 
reconstructed quantities as described in Sections~\ref{sec:fastsim} and \ref{sec:ultrafastsim}, respectively. 
In Sections~\ref{sec:turcal}, I will discuss how real data can be used to train the generative models, 
obtaining a fast simulation independent of, and therefore competitive with, the detailed simulation. 
Additional details on the training of the Generative Models for detector simulation are given in Sections~\ref{sec:optunapi}. Some strategies for the deployment of the models in the experiment 
software stack are reviewed in Section~\ref{sec:deployment}. 
Finally, I report on the comparison between reconstructed data and ultra-fast simulation
in Section~\ref{sec:validation}.
Section~\ref{sec:conclusion} concludes this paper with a short summary.

\section{The \emph{Detailed Simulation} of the LHCb Experiment}\label{sec:fullsim}
\begin{figure}
    \centering
    \includegraphics[width=\textwidth]{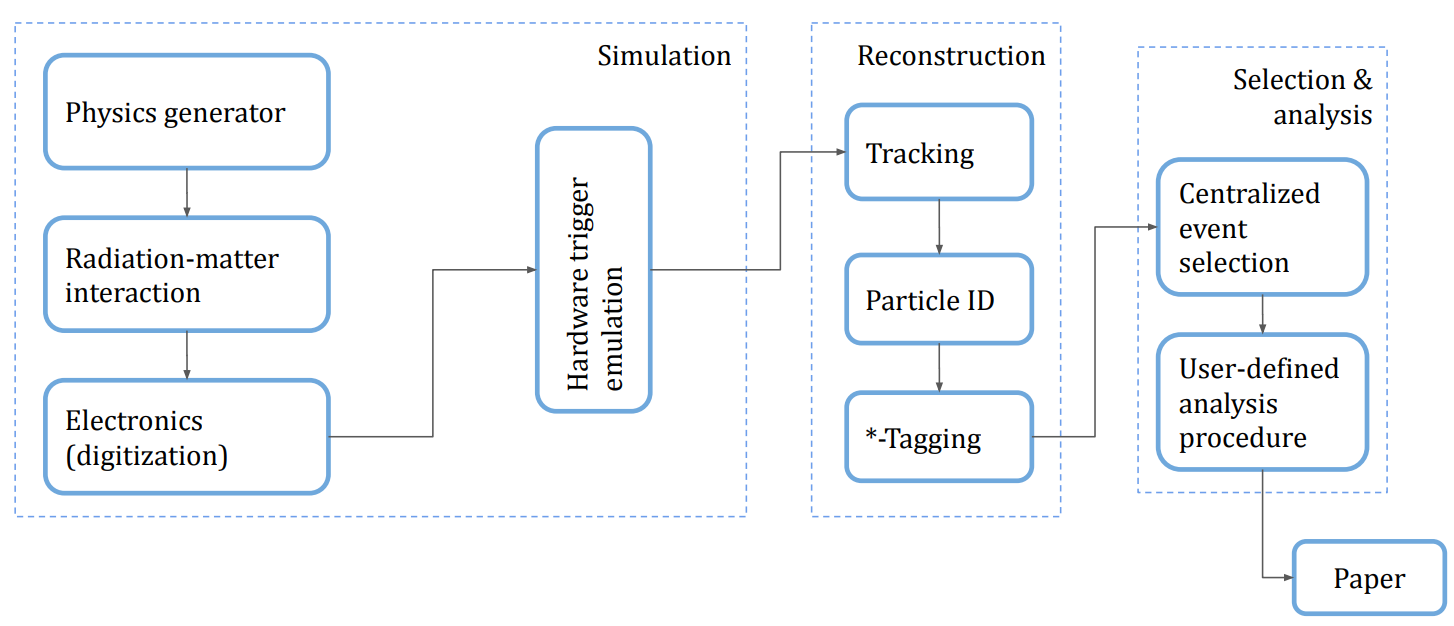}
    \caption{Flowchart of the Detailed Simulation processing}
    \label{fig:simflow}
\end{figure}

Detailed Simulation provides description based on the physical model of the detector response
to the incident particle.
%The most complete and safe option to obtain a data sample representative of the collision events
%recorded by the LHCb experiment is the \emph{Detailed Simulation}. 
As for most HEP experiments, the first step in the simulation of the LHCb experiment~\cite{LHCb:2011dpk}
is to simulate the proton-proton collisions with a 
Monte~Carlo generator such as PYTHIA~\cite{Sjostrand:2007gs}, EPOS~\cite{Pierog:2013ria}, 
or \textsc{BcVegPy}~\cite{Chang:2005hq}. The simulation of the heavy hadrons produced in the collision 
is obtained running \textsc{EvtGen}~\cite{Lange:2001uf}. The particles produced in primary collisions 
and in subsequent decays are then propagated through the detector and the energy deposited for 
radiation-matter interaction is computed with GEANT4~\cite{GEANT4:2002zbu}.
The energy deposits (or \emph{hits}) are then digitized~\cite{Corti:2006yx} 
and the hardware trigger is emulated~\cite{Aaij:2012me,LHCb:2018zdd} 
in order to provide to the event reconstruction pipeline 
an input as similar as possible to real data.
The reconstruction pipeline is composed of several algorithms dealing with 
\emph{tracking}, 
associating sets of aligned \emph{hits} to trajectories of charged particles (or \emph{track}s), 
the \emph{charged particle identification}, assigning to each track a likelihood (or other test statistics) 
for each mass hypothesis, and \emph{neutral object identification} mainly intended to identify 
photons impinging on the calorimeter. A final step in the reconstruction consists of algorithms 
considering sets of tracks to \emph{tag} some special property of the event. For example flavour tagging is 
a technique based on the statistical study of the particles produced in combination with a $b\bar b$ quark pair
developed to discriminate hadrons produced as $B^0$ or $\bar B^0$ and study flavour 
oscillations~\cite{LHCb:2016mtq,LHCb:2016yhi,LHCb:2015olj,LHCb:2012dgy}.

The reconstructed events are then processed with a centralised selection framework splitting the 
events in streams depending on the area of the LHCb physics programme. The streams are finally processed by 
the LHCb members individually to perform statistical analyses on sets of events and draw conclusions on 
the underlying physics model.

The procedure to obtain \emph{Detailed Simulation} samples is depicted as a flowchart in Figure~\ref{fig:simflow}.

\section{Models for the energy depositions in the calorimeters}\label{sec:fastsim}
The most common application of Machine Learning to Simulation is predicting the energy 
deposits of particles in the electromagnetic calorimeter. Indeed, the detailed simulation of 
the electromagnetic and hadronic showers in the calorimeters is computational espensive, 
accounting for more than half of the total CPU demand from the LHC experiments~\cite{Boccali:2019yhm,Bozzi:2020ogn}.
Intense research activity to speed-up the calorimeter simulation is attested by the 
many publications on the subject from the ATLAS~\cite{ATLAS:2021pzo, Krause:2021ilc}
and CMS~\cite{Paganini:2017dwg,Rahmat:2012fs,Abdullin:2011zz} Collaborations, but also in the 
context of the effort towards GeantV~\cite{Amadio:2018tnh, Khattak:2021ndw}. A software package 
dedicated to the parametric simulation of the calorimeters of ATLAS, CMS and experiments at future colliders
is also being developed under the name of Delphes~\cite{deFavereau:2013fsa}.

Both traditional parametrization techniques and methods based on Machine Learning have been explored. 
When machine learning is employed, it often relies on 
Deep Convolutional Generative Adversarial Neural Networks  (DCGANs)~\cite{radford2016unsupervised} 
to predict the energy deposited in each active volume (voxel) by taking into account the correlation between 
adjacent regions. For example, the model developed for the LHCb experiment solutions based on 
conditional DCGANs~\cite{Chekalina:2018hxi} and a combination of conditional DCGANs and Variational Autoencoders
(VAEs)~\cite{Sergeev:2021adf}. In both models the idea is to take as an input the parameters of each particle
(either a charged one or a photon) such as its momentum and slopes, and translate them to a randomly 
generated  \emph{image} of the shower in the electromagnetic calorimeter. 
The procedure can then be repeated for each track in 
the event to reconstruct the overall response of the calorimeter.

These methods provide an output which is intended to be processed with the whole reconstruction pipeline.
The resulting simulated samples contain all the features exactly as the detailed simulation, 
including, for example, the tagging information based on collective features.
On the other hand, achieving good representation of the reconstructed quantities is challenging because 
the reconstruction algorithms are not known to the training algorithms, and therefore 
the reproduction of reconstructed quantities can be optimised
only indirectly. 

\section{Models for the reconstructed quantities}\label{sec:ultrafastsim}
To achieve a further speed-up of the Simulation improving the quality of the reconstructed 
quantities, Musella and Pandolfi proposed to use GANs to simulate directly the variables obtained 
from the reconstruction pipeline~\cite{Musella:2018rdi}. 
This approach allows for a faster and more accurate simulation with respect to the simulation of 
the energy deposits, but the multi-particle effects on the reconstruction becomes more difficult 
to take into account because the reconstruction algorithms are parametrise themselves. 
As a consequence, the high-level physics quantities obtained through the \emph{tagging} 
part of the reconstruction pipeline represent an additional challenge
in this so called \emph{ultra-fast simulation}. 

Nonetheless, the method had success enabling fast simulation of detectors beyond 
calorimeters, providing for example the output of the LHCb RICH detectors~\cite{Maevskiy:2019vwj} 
and muon system~\cite{Sassoli:bachelor}.
In both cases, a conditional GAN has been trained to reproduce the PID-related observables 
(or, more technically, the logarithmic 
likelihood ratio between the muon, kaon, or proton hypothesis and the pion hypothesis).
The reconstruction algorithm used to obtain these quantities is rather complex, involving a event-wise 
modelling of the energy deposits in the calorimeter under various combinations for the mass 
hypotheses~\cite{LHCbRICHGroup:2012mgd}, but the effect of 
inter-particle correlations on the likelihoods, 
can be effectively modelled by including the detector occupancy among the GAN \emph{conditions} together 
with track momentum and pseudorapidity. 

\section{Training on simulated or real data}\label{sec:turcal}
The most common procedure to train Generative Models to parametrise the detector response
is to generate samples through the Detailed Simulation pipeline and then train the 
model to associate generator-level variables to reconstructed quantities. 
This simple technique, however, is intrinsically limited to the quality of the Detailed Simulation or, 
in other words, a fast simulation obtained training on detailed simulation will be at most as accurate
as detailed simulation. 

As an interesting alternative, the Generative Models can be trained on real data if following two conditions are 
satisfied.
\begin{itemize}
    \item \emph{The acquired dataset is unbiased.} The training data should be selected without applying
        selection criteria that, directly or indirectly, modify the distributions of the reconstructed 
        quantities beyond the effect taken into account by the \emph{conditions} passed as inputs 
        to the Generative Model (such as the kinematics or the detector occupancy).
        For example, a very common bias is introduced by trigger algorithms designed to select muons. 
        A training sample obtained without excluding candidates selected by these lines will result into
        Generative Models overestimating the probability of misidentification of hadrons as muons, because 
        the training sample was enriched of hadrons with a detector response more similar to that of a muon than 
        the average hadron. In LHCb this is avoided providing dedicated trigger lines and a dedicated 
        data processing scheme for the calibration data used to train the Generative Models for the
        ultra-fast simulation~\cite{Aaij:2018vrk}.
        
    \item \emph{The generative model supports background subtraction.}
        Calibration samples are usually as abundant and pure as possible, but in order to avoid selection bias,
        the selections are often loose and some background contribution is expected. 
        For all the models with a loss function derived more or less directly 
        from the maximum likelihood principle, it was demonstrated that weighting the events with 
        properly-computed weights the effect of the background component on the trained network is 
        null~\cite{Xie:2009rka, Borisyak:2019vbz}. These background-subtracting weights, sometimes called 
        $_s\mathcal W$eights in the context of High Energy Physics, are computed taking into account 
        the models for the signal and the background components in a discriminant variable (\emph{e.g.}
        the invariant mass of the combination of reconstructed particles) assumed to be independent of 
        all the other (target) variables involved in the background subtraction~\cite{Pivk:2004ty}.
\end{itemize}

It is worth noticing that an ultra-fast simulation trained on real data is subject to systematic 
errors in the description of the detector response which are completely independent of the detailed simulation.
Detailed and ultra-fast simulation can then cross-check each other, while providing a measure 
of the simulation accuracy depending on the specific phase space relevant to each analysis. 

\section{Automatic hyperparameter optimisation for generative models}\label{sec:optunapi}
The training of GANs is well known within the Machine Learning community to be a difficult 
task~\cite{10.5555/2969033.2969125}. 
Reaching the Nash equilibrium between the Generator and the Discrminator requires 
careful tuning of the hyperparameters, such as the learning rate or the network structure. 
Unfortunately, the absolute value of the loss function used to train adversarial models 
is not meaningful: a discriminator that finds more difficult to separate real and fake
candidates may mean that the generator is better (and so is the accuracy of the generated sample) 
or that the disciminator is worse (and so is the accuracy of the generated sample). 

In order to be able to use Bayesian hyperparameter optimisation frameworks,
such as \texttt{optuna}~\cite{akiba2019optuna} or 
\texttt{scikit-optimize}~\cite{head_tim_2020_4014775}, LHCb adopted a \emph{goodness-of-fit} method
based on the ability of a BDT classifier to distinguish training and generated data 
entries~\cite{Williams:2010vh, Weisser:2016cnc}. 
The BDT classifier is a third player, independent of the adversarial 
training procedure, and is based on a different classification method with different
underlying hypotheses. To provide an estimate of the uncertainty on quality metric 
obtained from the BDT, the same algorithm is trained several times on the same sample
with different random seed. The average and its uncertainty are used as guidance for 
optimising the hyperparameters using a distributed version of \texttt{optuna}, 
named OptunAPI~\cite{matteo_barbetti_2021_5538989}. 
With this strategy, it is possible to automate the training of GANs on multiple GPU 
nodes making the production of a different model per datataking period accessible. 

\section{Deployment in the software stacks of the experiments}\label{sec:deployment}
The training and the validation of the generative models relies on Python frameworks
such as \emph{scikit-learn}~\cite{scikit-learn} for the pre- and post-processing
stages and either 
\emph{tensorflow}~\cite{tensorflow}
or \emph{PyTorch}~\cite{NEURIPS2019_9015} 
for the description of the neural network model. 

The pipeline built in Python must then be deployed in the software stack of the experiments,
which is most often written in C++ and released few times a year. 
Both \tensorflow and PyTorch provide C/C++ APIs to deploy neural network models, however 
the build system of these large packages often conflicts with the build system of the 
experiment's software. The same applies for the multi-thread schedulers, designed to meet
the high-throughput requirements of LHC experiments on one side and to speed-up the 
evaluation of the networks on the other. 

In the context of the CMS and LHCb collaborations, \tensorflow has been succesfully 
integrated within the experiment software%
\footnote{
see \href{https://gitlab.cern.ch/mrieger/CMSSW-DNN}{CMSSW-DNN} 
and \href{https://gitlab.cern.ch/lhcb/LHCb/-/tree/master/Tools/GaudiTensorFlow}{GaudiTensorFlow}
on GitLab.
} 
providing evaluations of arbitrarily complicated networks with short latency.
A limitation derived from this approach is that all the networks within the experiment software 
must share the same version of \tensorflow, which makes the maintenance effort rather 
significant. An intestersting alternative is to provide an inference service external to the
experiment software, possibly running on the same machine and interfaced to the data processing software 
framework via REST APIs~\cite{Kuznetsov:2020mcj}. This approach effectively decouples 
production code and the 
internal architecture of the server which may be based on whatever framework with whatever 
thread scheduler and hardware acceleration as long as it is consistent with the REST interface. 
Several servers can run concurrently to rely on multiple frameworks or framework versions. 
Due to the large latency, however, this technique is more suitable to process large events 
with thousands of tracks than to produce ultra-fast simulation, where the time spent per event is 
of the same order as the REST API latency. 

To reduce the latency as much as possible, the ATLAS Collaboration has been developing 
LWTNN~\cite{daniel_hay_guest_2021_5082190}, a C++-compiled generic interface for Deep
Neural Networks. LWTNN is composed of an engine compiled together with the experiment 
software and released twice a year, and reads configuration files describing the network 
architecture and its weights, to reconstruct the model at runtime. 
To make the evaluation even faster, LHCb has been developing RNNGenerator\footnote{see
\href{https://gitlab.cern.ch/vjevtic/rnngenerator}{RNNGenerator} on GitLab}
to produce C++ code for the Neural Network and compile it together 
with the experiment's software without any dependence on external frameworks. 
On the same line, \scikinC\footnote{
see \href{https://github.com/landerlini/scikinC}{\scikinC} on GitHub
} is a package designed to convert \texttt{scikit-learn} models in C and distribute them 
as optimised single-thread shared objects. More recently, support to simple Keras~\cite{chollet2015keras} 
feed-forward models was added. Hence, \scikinC is much more limited than LWTNN and 
RNNGenertor in terms of supported architectures, but eliminates almost any overhead and 
enables the distribution of models, including architecture and weights, as data objects.

The ultra-fast simulation framework of the LHCb Collaboration was originally designed 
to rely on GaudiTensorFlow, but an implementation based on \scikinC was found 
significantly faster. GaudiTensorFlow is now being phased out and the ultra-fast simulation 
migrated to \scikinC. 

\section{Validation}\label{sec:validation}

\begin{figure}
    \centering
    \begin{minipage}{\textwidth}
        \includegraphics[width=0.5\textwidth]{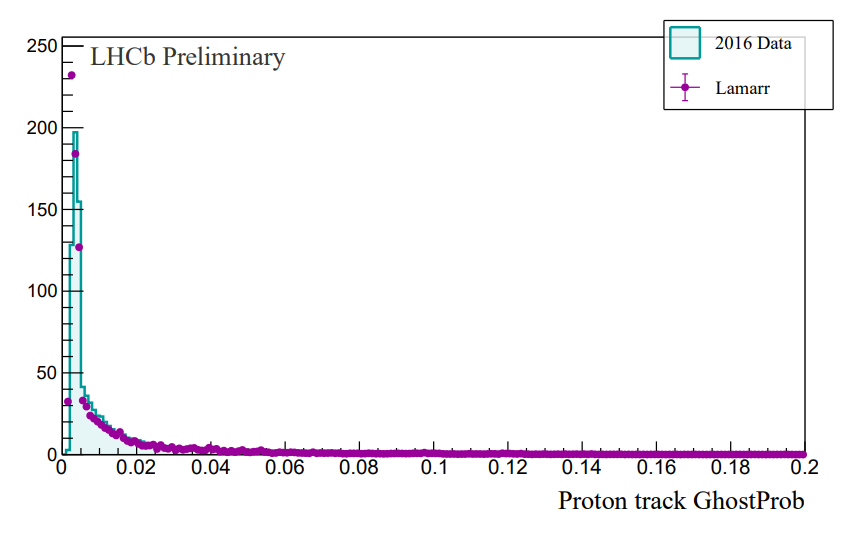}
        \includegraphics[width=0.5\textwidth]{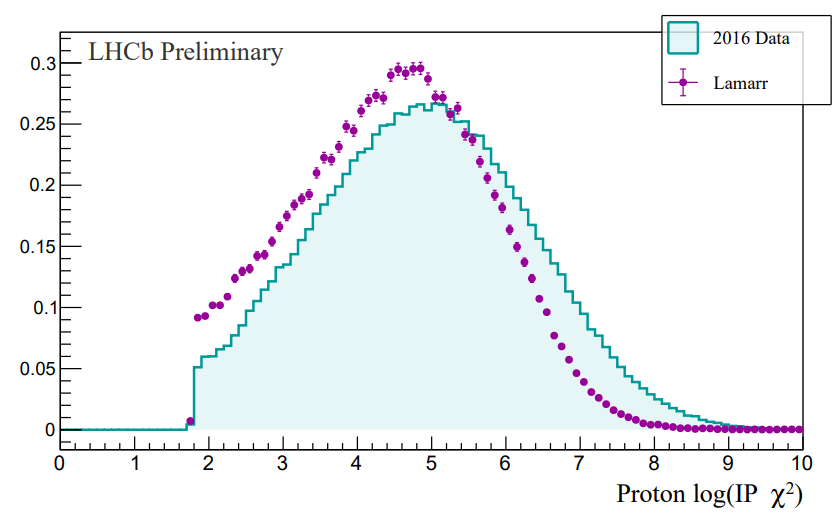}
    \caption{\label{fig:trackingvar} 
        Comparison of the tracking-related variables for protons obtained from 
        tagged $\Lambda_c^+ \to p K^- \pi^+$ decays in calibration data
        and through the whole ultra-fast simulation pipeline as implemented in Lamarr.
        Figure reproduced from Ref.~\cite{LHCB-FIGURE-2019-017}.
    } 
    \end{minipage}
    \begin{minipage}{\textwidth}
        \includegraphics[width=0.5\textwidth]{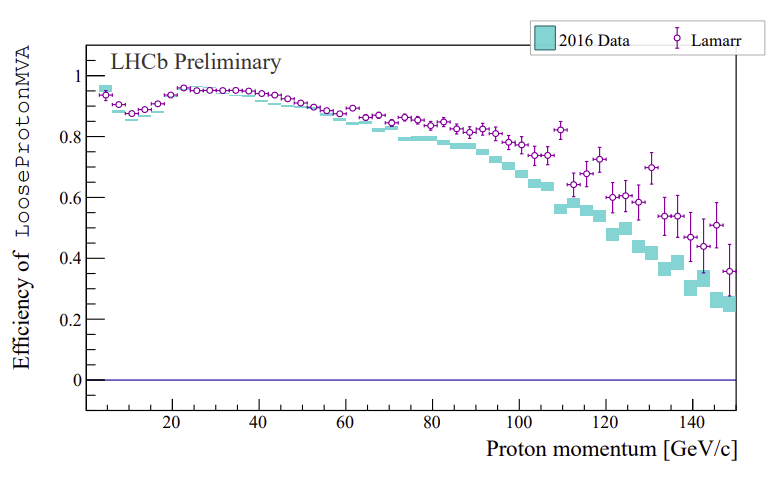}
        \includegraphics[width=0.5\textwidth]{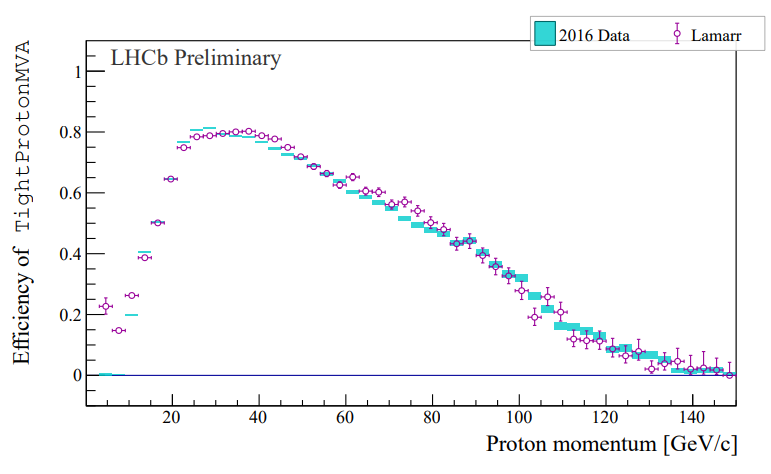}
    \caption{\label{fig:pidvar} 
        Comparison of the efficiency of two PID requirements with different thresholds
        as obtained from
        tagged $\Lambda_c^+ \to p K^- \pi^+$ decays in calibration data from Lamarr. 
        Figure reproduced from Ref.~\cite{LHCB-FIGURE-2019-017}.
    } 
    \end{minipage}
\end{figure}

The fast-simulation validation campaign is performed comparing data from a 
calibration sample with the results from Lamarr, the ultra-fast simulation package 
under development within LHCb. 
To validate the generalisation properties of the models, different decay models 
are chosen for the training and the validation of the simulation pipeline. 
For example, the proton identification is trained using tagged protons from 
$\Lambda^0 \to p\pi^-$ decays while the validation relies on tagged protons
from $\Lambda_c^+ \to p K^- \pi^+$ decays with the charm hadron produced in the
semileptonic decay $\Lambda_b^0 \to \Lambda_c^+ \mu^- X$. 

Figure~\ref{fig:trackingvar} reports the comparison between the calibration data
and the simulation for the response of a neural network classifier trained to 
identify ghost tracks and for the logarithm of the impact-parameter $\chi^2$, 
measuring the consistency of the proton track with the primary vertex.  
The former is a variable directly obtained as an output of a Generative Model,
while the latter is computed from the simulated properties of the track and 
primary vertex, and their (simulated) uncertainties.  

Figure~\ref{fig:pidvar} shows instead the selection efficiency of a criterion based 
on the response of classifier trained to distinguish the particle mass combining 
the response of the whole LHCb Particle Identification apparatus: two RICH detectors,
four Muon stations and the electromagnetic calorimeter. 
In this case of the simulation, the response of the classifier is obtained 
directly as an output of the generative model.
The selection efficiency for two different thresholds are shown as a function of the 
proton momentum, demonstrating the capability of the Generative Model to properly 
model the dependence of the detector response on the kinematics of the impinging particle. 
Similar plots are produced to check good representation of the dependence on the 
detector occupancy and other \emph{conditions}.  
For a deeper discussion of these and additional comparison plots, 
see Ref. \cite{LHCB-FIGURE-2019-017}.

\section{Conclusion}\label{sec:conclusion}
The simulation of the LHCb experiment is an expensive tasks posing severe pressure
on the pledged computing resources. 
Several solutions to speed-up the simulation are being explored, among those 
Machine-Learning Generative Models are extremely promising. 
Solutions have been proposed to model both the energy depositions in the calorimeters
and the analysis-level quantities features, presenting complementary advantages. 

The techniques used to train the model on calibration data enables to go beyond 
an approximation of the full simulation, obtaining an ultra-fast simulation 
potentially competitive with the full simulation and providing useful cross-checks. 

Both the training and the deployment of Generative Models are active research 
subjects to keep improving the quality of the simulated features. I discussed 
as an example the Bayesian hyperparameter optimisation based on OptunAPI, and 
several deployment strategies with different balance between performance and support 
to neural network models. 

Lamarr, the ultra-fast simulation framework under development in LHCb, combines
several traditional parametrisations and Generative Models to produce simulated output 
as similar as possible to reconstructed data. The comparison of the simulated 
quantities with calibration samples not used for the training confirms good 
generalisation properties. 

Upcoming releases of Lamarr will improve the quality of the simulation and extend 
the reach of the simulation by adding several reconstructed quantities, 
providing a powerful tool within the wide Simulation palette of the LHCb experiment.


\bibliographystyle{JHEP}
\bibliography{main}
\addcontentsline{toc}{section}{References}

\end{document}